# *Transscleral Micropulse Diode Laser for Treatment of Diabetic Retinopathy*

Hamed Nasreldin Taha[1*], Ahmed Gamal[2*], Sarah Hamed Nasreldin[3*]

[1] *Faculty of Medicine, Al Azhar University, Cairo, Egypt*
[2] *Saudi German Hospital, Jeddah, Saudi Arabia*
[3] *Faculty of Medicine, Cairo University*

## Abstract

In this study, we introduced a new simple effective technique for the treatment of diabetic retinopathy by the treatment of the periphery of the retina by transscleral micropulse diode laser (Taha Technique). Macular edema was improved and vitreous hemorrhage was disappeared in most of the cases.

## Introduction/Background:

Micropulse Diode Laser Photocoagulation **(MD)** is designed to target The Retinal Pigment Epithelium **(RPE)** melanocytes while avoiding photoreceptors damage. The term "subthreshold" refers to photocoagulation that does not produce visible intraretinal damage or visible scarring either during or after treatment. It is well known that burns cannot be detected by clinical examination or by Intravenous Fluorescein Angiography **(IVFA)** and Fundus Autofluorescene (FAF) [1,2].

Photocoagulation of all layers of retina is not needed to get the therapeutic effect on the retina by laser. The RPE is essential for repairing the outer and inner blood-retinal barrier regardless of the type and location of laser application. The absence of micro-pulse chorioretinal laser damage permit retreatment of the same area. The absence of chorioretinal scaring allows for an overlapping application of burns. Frequent retreatment of involved retinal areas is also possible without fear of creating confluent retinal scarring [3].

When the threshold of sublethal cellular injury is reached via the cumulative addition of denatured proteins, transcriptional activation of cytokine expression, release of growth factors and up regulation of matrix metalloproteinases occurs. It has the ability to up regulate various biochemical mediators with anti-angiogenic activity, such as Pigment Epithelium Derived Growth Factor (PEDGF). Also, the treatment by micro-pulse laser stimulates the release of factors that increase angiotensin II and increase receptor activity, enabling inhibition of Vascular Endothelial Growth Factor (VEGF).The reduction in VEGF also reduces vascular permeability [4,5].

## Patients & Methods:

This study included 46 eyes of 34 patients who had diabetic macular edema and/or proliferative diabetic retinopathy. The Patients in this study were divided into 2 groups: Group (1) has macular edema thickness less than 400um. Group (2) has proliferative diabetic retinopathy with vitreous hemorrhage. Transscleral Micropulse Diode Laser treatment was given to the periphery of the retina. Laser sessions ranged from 1 to 3 in the first group and up to 5 sessions in the second group. The same laser parameters were used for each patient. Only the number of laser shots varied between patients. Pre- laser treatment and in follow up, visual acuity , OCT, colored fundus photo and fluorescein angiography were done.

## Results:

Patients follow up ranged from one month to one year. Visual acuity were improved in 68% of treated eyes in group 1 , and in 84 % of eyes in group 2.

Statistically significant reduction of macular edema was in 70 % of eyes in group 1. Total disappearance of vitreous hemorrhage and reduction of diabetic retinopathy was observed in 80 % of eyes in group 2. No adverse laser events occurred, no laser lesions were detectable clinically or angiographically after treatment.

**Group (1):**

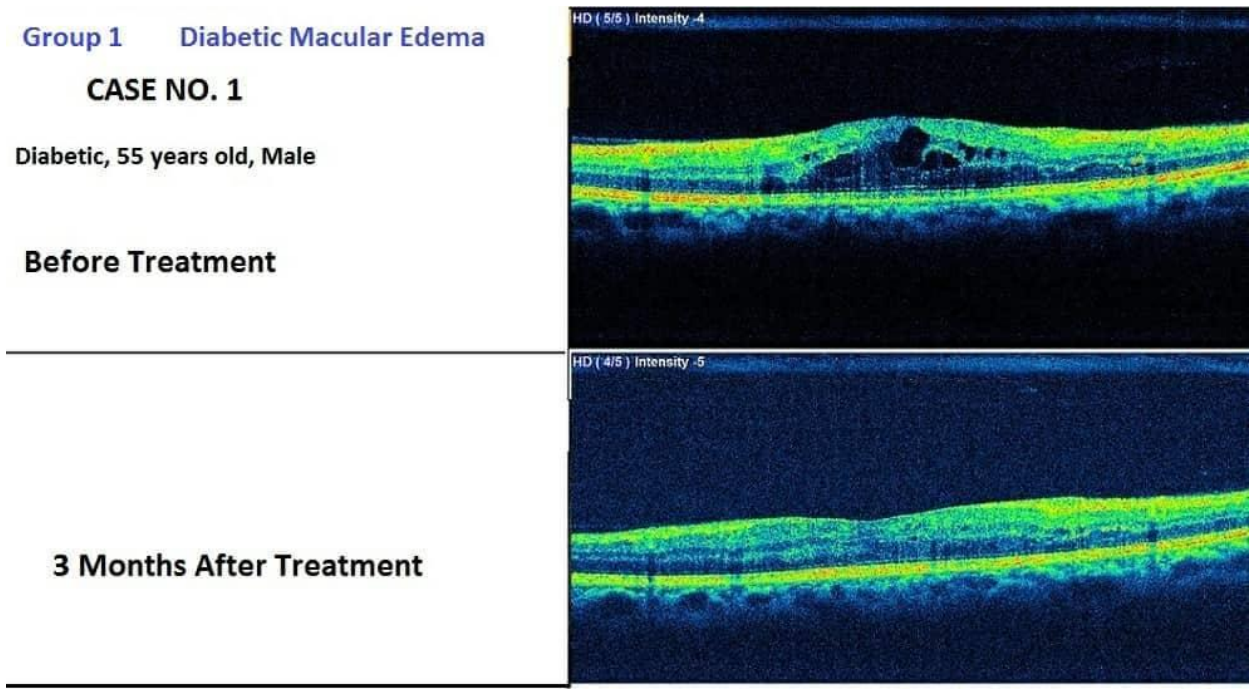

**Figure (1):** Case number 1 with diabetic macular edema which shows marked improvement within 3 months after Micropulse Diode Laser Photocoagulation.

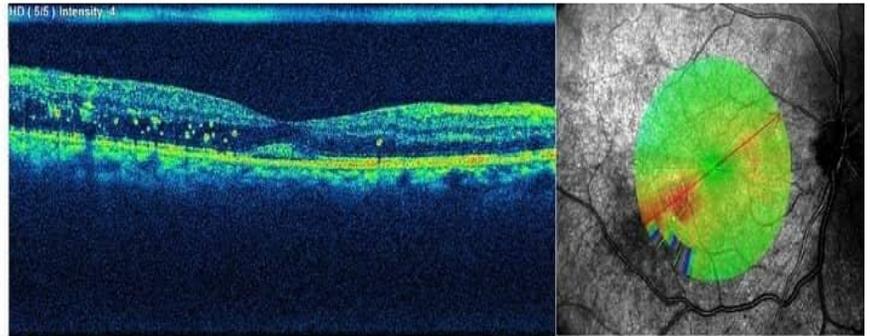
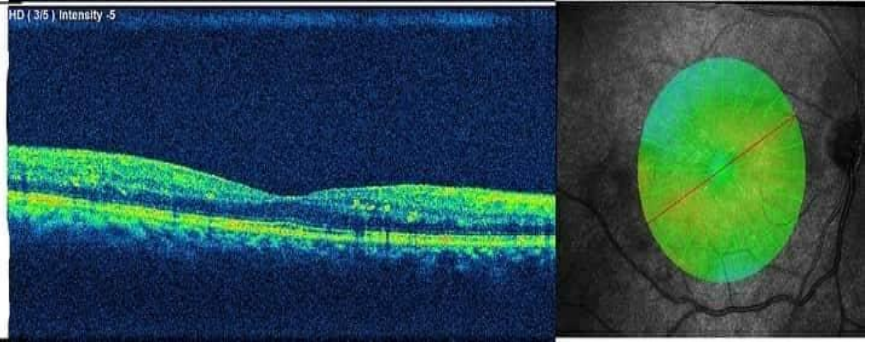

**Figure (2):** Case number 2 with diabetic macular edema which shows marked improvement within 3 months after Micropulse Diode Laser Photocoagulation.

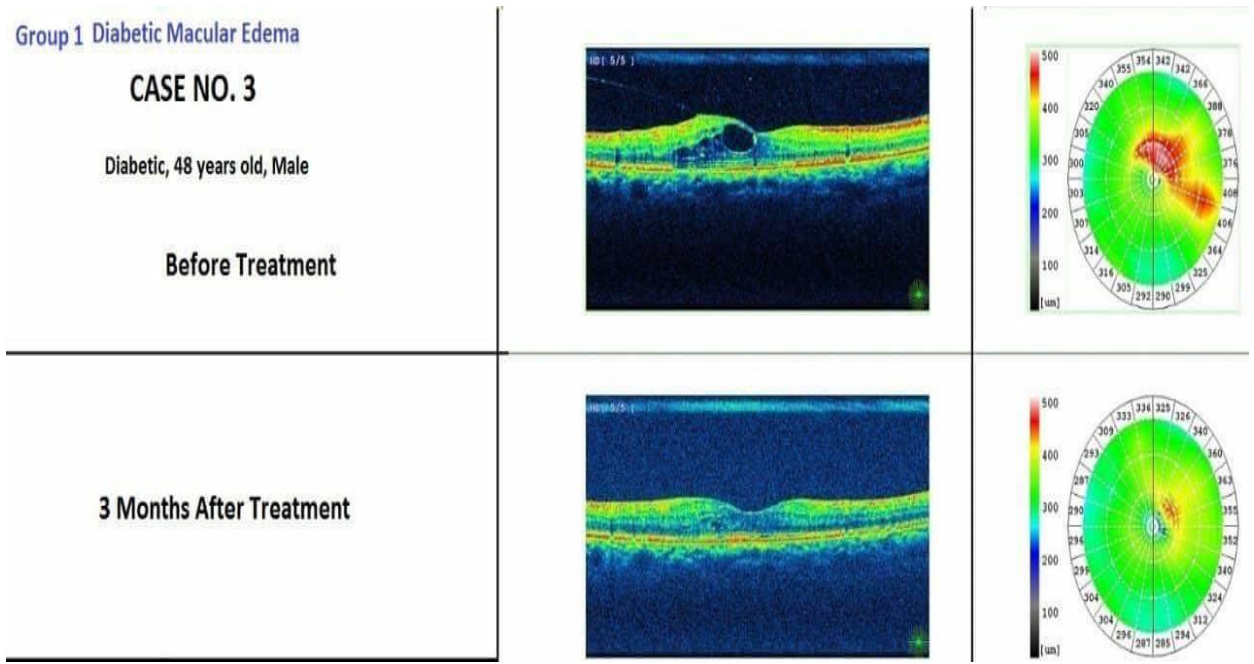

**Figure (3):** Case number 3 with diabetic macular edema which shows marked improvement within 3 months after Micropulse Diode Laser Photocoagulation.

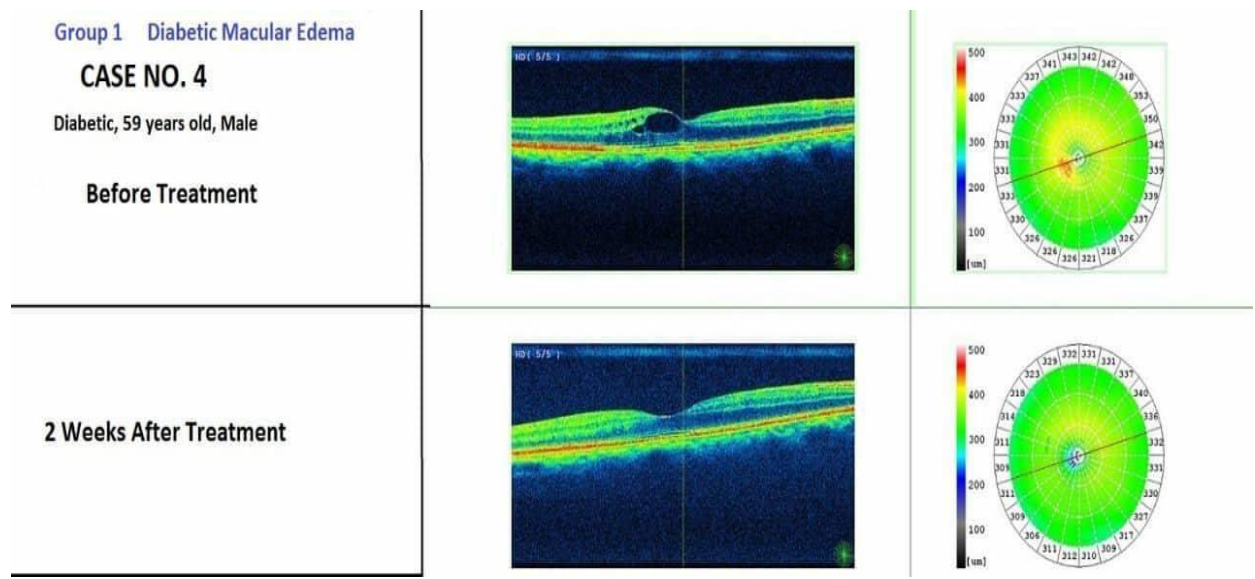

**Figure (4):** Case number 4 with diabetic macular edema which shows marked improvement within 2 weeks after Micropulse Diode Laser Photocoagulation.

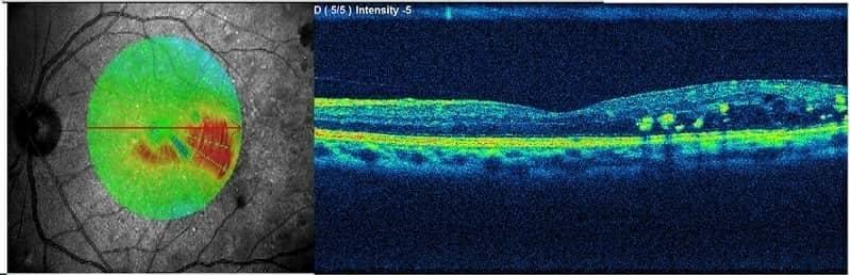
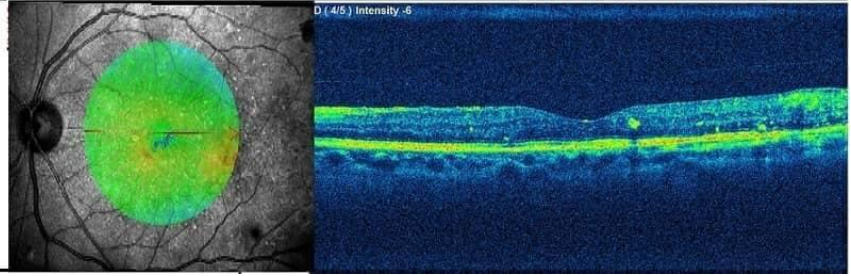

**Figure (5):** Case number 5 with diabetic macular edema which shows marked improvement within 5 months after Micropulse Diode Laser Photocoagulation.

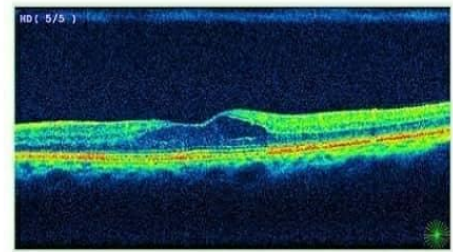
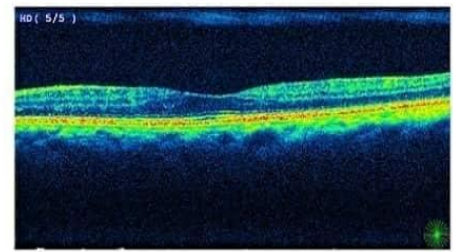

**Figure (6):** Case number 6 with diabetic macular edema which shows marked improvement within 3 months after Micropulse Diode Laser Photocoagulation.

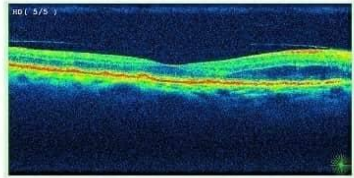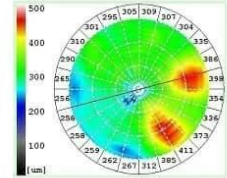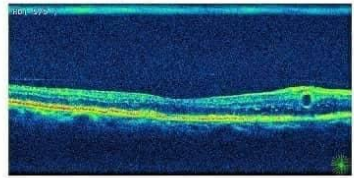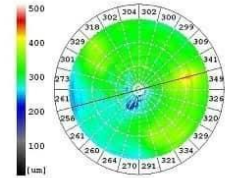

**Figure (7):** Case number 7 with diabetic macular edema which shows marked improvement within 3 months after Micropulse Diode Laser Photocoagulation.

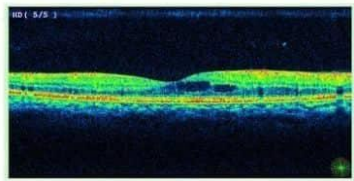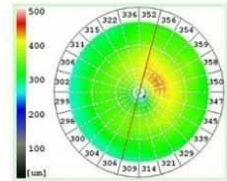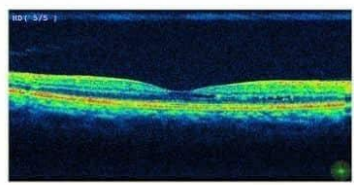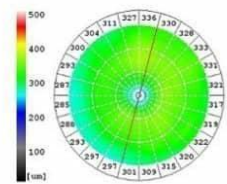

**Figure (8):** Case number 8 with diabetic macular edema which shows marked improvement within 2 months after Micropulse Diode Laser Photocoagulation.

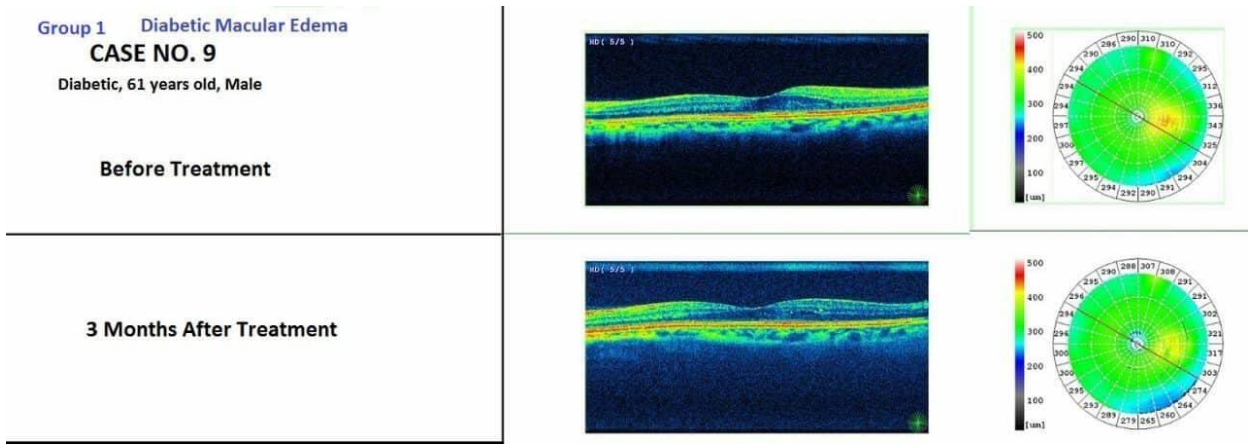

**Figure (9):** Case number 9 with diabetic macular edema which shows marked improvement within 3 months after Micropulse Diode Laser Photocoagulation.

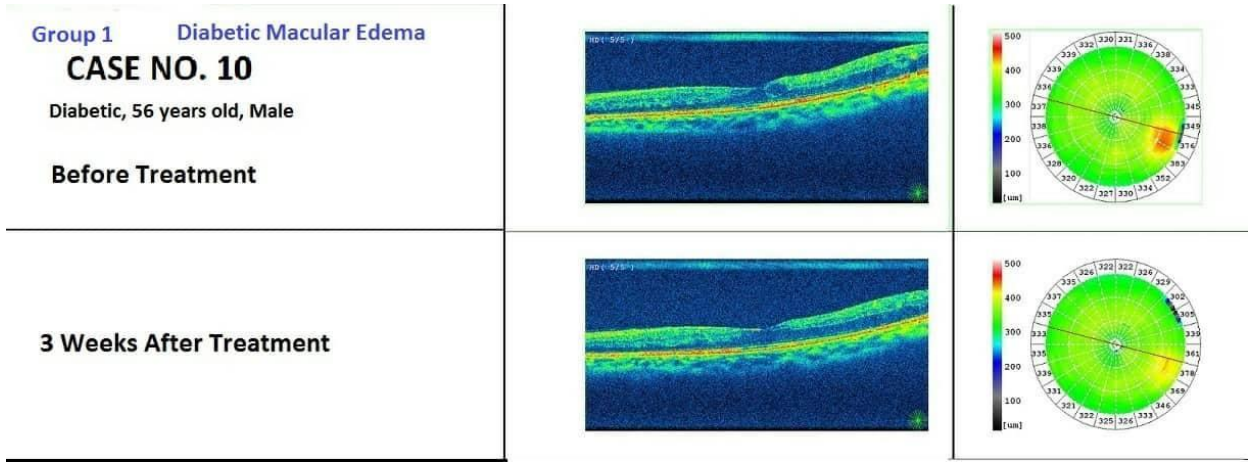

**Figure (10):** Case number 10 with diabetic macular edema which shows marked improvement within 3 weeks after Micropulse Diode Laser Photocoagulation.

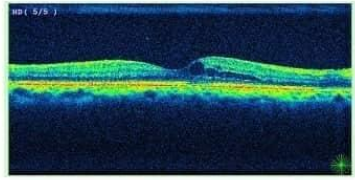
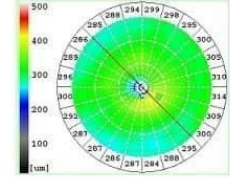
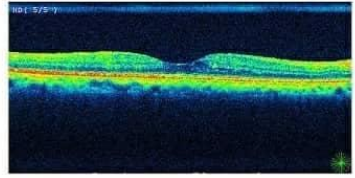
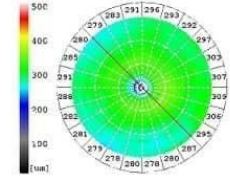

**Figure (11):** Case number 11 with diabetic macular edema which shows marked improvement within 3 months after Micropulse Diode Laser Photocoagulation.

**Group (2):**

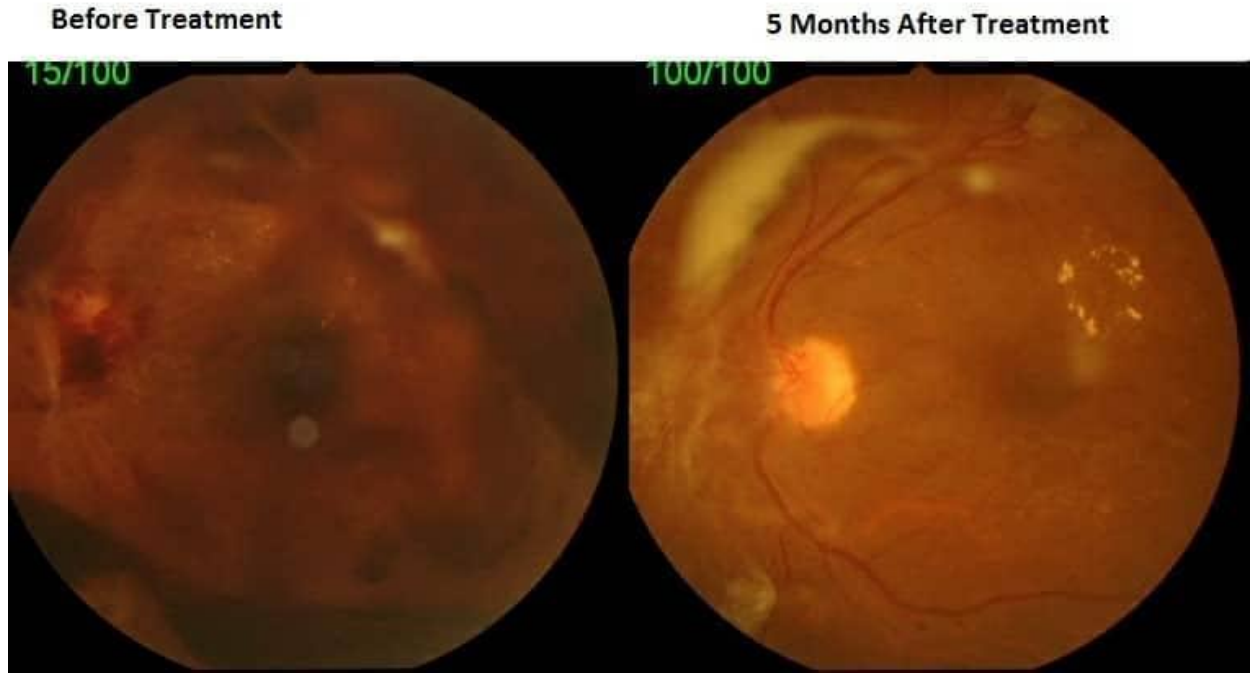

**Figure (12):** Case number 12 with proliferative diabetic retinopathy and vitrous heamorrhage which shows marked improvement within 5 months after Micropulse Diode Laser Photocoagulation.

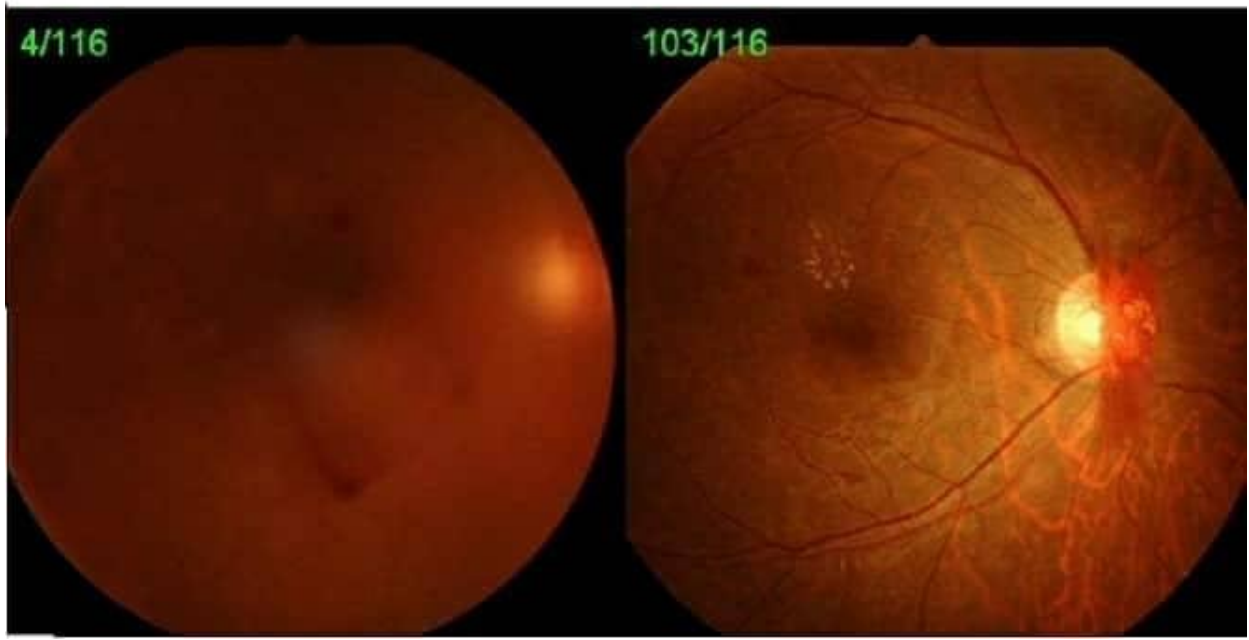

**Figure (13):** Case number 13 with proliferative diabetic retinopathy and vitrous heamorrhage which shows marked improvement within 2 months after Micropulse Diode Laser Photocoagulation.

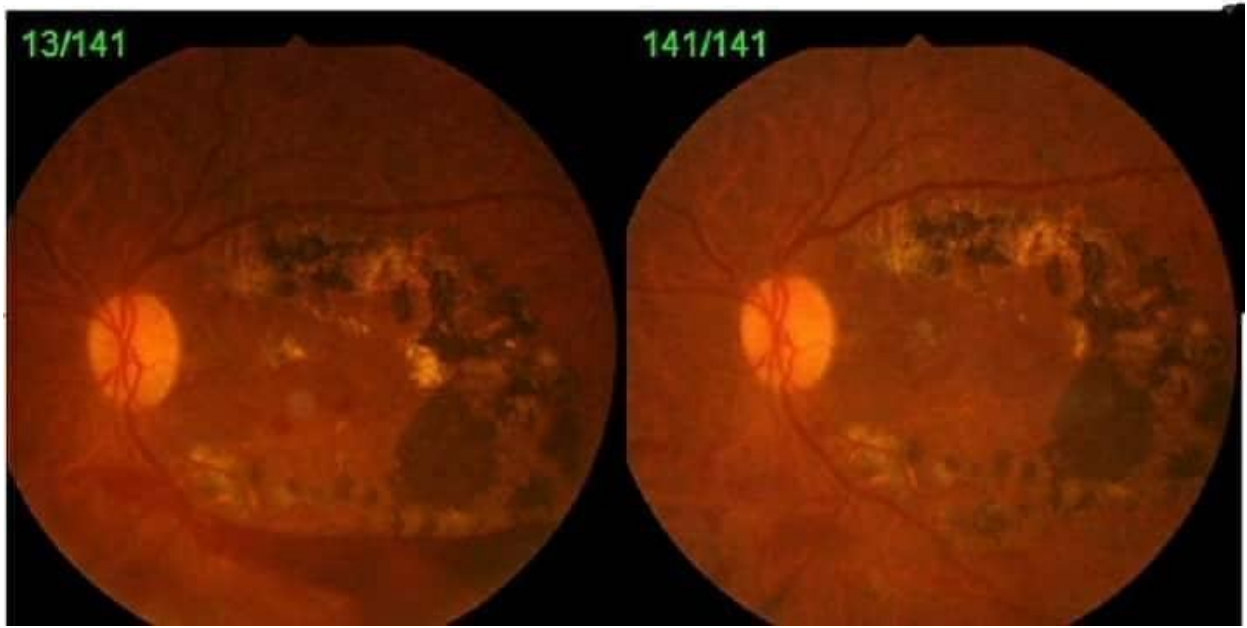

**Figure (14):** Case number 14 with proliferative diabetic retinopathy and vitrous heamorrhage which shows marked improvement within 5 months after Micropulse Diode Laser Photocoagulation.

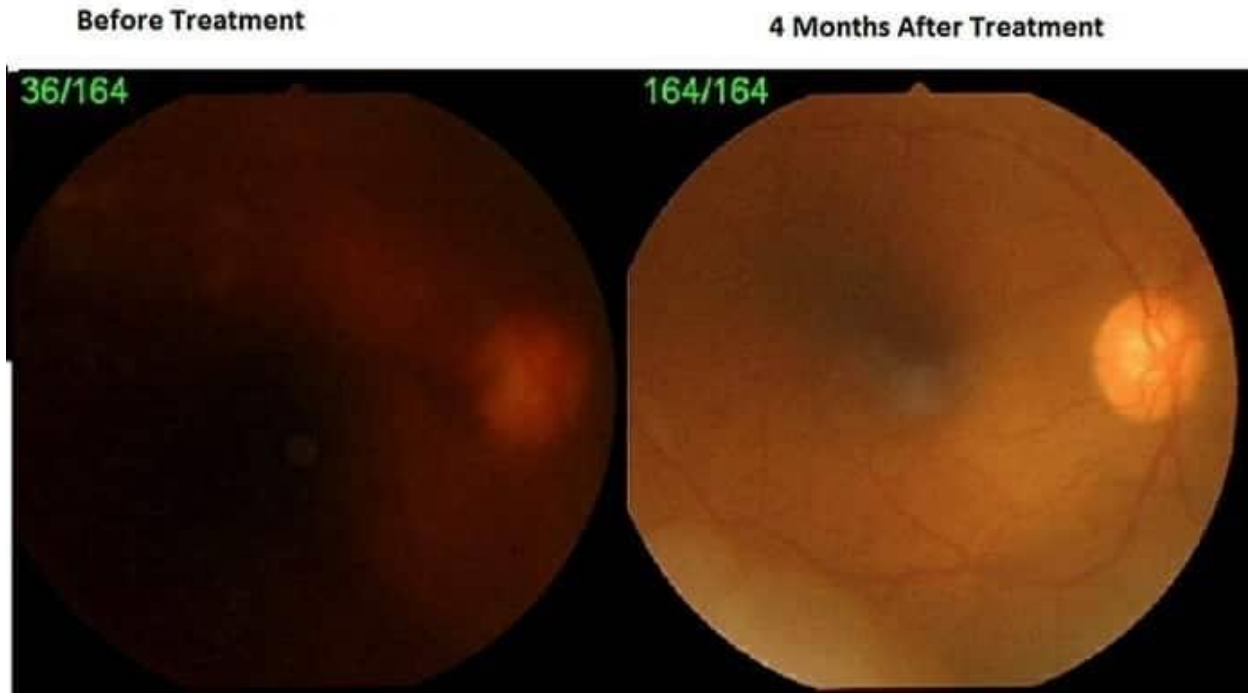

**Figure (15):** Case number 15 with proliferative diabetic retinopathy and vitrous heamorrhage which shows marked improvement within 4 months after Micropulse Diode Laser Photocoagulation.

## Conclusions:

Transscleral micropulse diode laser treatment of diabetic macular edema and proliferative diabetic retinopathy (Taha Technique) is an easy new technique, effective and safe treatment way.

# References

1. Sivaprasad, S., Elagouz, M., McHugh, D., Shona, O. & Dorin, G. Micropulsed Diode Laser Therapy: Evolution and Clinical Applications. *Surv. Ophthalmol.* **55,** 516–530 (2010).

2. Kiire, C., Sivaprasad, S. & Chong, V. Subthreshold micropulse laser therapy for retinal disorders. *Retin. Today* **1,** 67–70 (2011).

3. Palanker, D. Evolution of Concepts and Technologies in Ophthalmic Laser Therapy. *Annu. Rev. Vis. Sci.* **2,** 295–319 (2016).

4. Ogata, N., Tombran-Tink, J., Jo, N., Mrazek, D. & Matsumura, M. Upregulation of pigment epithelium-derived factor after laser photocoagulation. *Am. J. Ophthalmol.* **132,** 427–429 (2001).

5. Flaxel, C., Bradel, J., Acott, T. & Samples, J. R. Retinal Pigment Epithelium Produces Matrix Metalloproteinases After Laser Treatment. *Retina* **27,** 629–634 (2007).